\documentclass[aps,prl,twocolumn,superscriptaddress,showpacs,nofootinbib]{revtex4}
\usepackage{times}
\usepackage[dvips]{graphics,graphicx}
\usepackage{amsmath, amssymb}
\begin{document}
\title{Mass Shift and Width Broadening of $J/\psi$ in hot gluonic plasma
from QCD Sum Rules}
\author{Kenji Morita}
\email{morita@phya.yonsei.ac.kr}
\author{Su Houng Lee}
\email{suhoung@phya.yonsei.ac.kr}
\affiliation{Institute of Physics and Applied Physics, Yonsei
University, Seoul 120-749, Korea}
\date{\today}
\begin{abstract}
 We investigate possible mass shift and width broadening of $J/\psi$ in
 hot gluonic matter using QCD sum rule. Input values of gluon condensates at
 finite temperature are extracted from lattice QCD data for the energy
 density and pressure. Although stability of the moment ratio is
 achieved only up to
 $T/T_{\text{c}}\simeq 1.05$, the gluon condensates cause a decrease of the
 moment ratio, which results in change of spectral
 properties. Using the Breit-Wigner form for the phenomenological side,
 we find that mass shift of $J/\psi$ just above $T_{\text{c}}$ can reach
 maximally
 200 MeV and width can broaden to dozens of MeV.
\end{abstract}
\pacs{12.38.Mh,11.55.Hx,14.40.Gx}

\maketitle

Heavy quarkonia have been regarded as a very useful probe of quark-gluon
plasma (QGP) which can be created in relativistic heavy ion
collisions. Since Matsui and Satz argued
\cite{Matsui_PLB178} that the $J/\psi$ suppression could be a signature of
QGP formation in heavy ion collisions, extensive works have been
performed on the subject both experimentally
\cite{PHENIX_Jpsilatest} and theoretically
\cite{Kharzeev_QM06}. However, in contrast to early expectations, recent
lattice calculations
suggest $J/\psi$ can survive up to at least $T\sim 1.5T_{\text{c}}$
\cite{Umeda_IJMP16,Asakawa_PRL92,Datta_PRD69}. Hence, change of spectral
properties, which cannot be seen within current resolution of lattice
calculations, may exist in QGP and can reflect the properties of the strongly
coupled QGP, at temperature not so higher than $T_{\text{c}}$. One
should note, however, that in such a temperature region, the system 
is highly non-perturbative one \cite{Blaizot_QM06}, and any analyses should
consistently treat the non-perturbative aspects of QCD. In this respect,
QCD sum rule is a suitable approach that can be used for analyses of
hadron properties at these temperatures.

In the present paper, we investigate the behavior of $J/\psi$ in gluonic
plasma
slightly above $T_{\text{c}}$
using QCD sum rules \cite{Shifman_NPB147}. The QCD sum
rule has been extensively used for studying in-medium
properties of both light and heavy hadrons as a reliable and
well-establish method. 
For heavy quark systems, it is more reliable
because all relevant condensates are known and their temperature
dependence are easily extracted from
lattice QCD.
Here we follow the method used in analyses of $J/\psi$ in
vacuum \cite{Reinders_NPB186} and in nuclear matter
\cite{Klingl_PRL82,hayashigaki99:_jpsi} and study $J/\psi$ at finite
temperature within the quenched approximation.

We start with the time ordered current-current correlation function
\begin{align}
(q^\mu q^\nu -q^2 g^{\mu\nu})\tilde{\Pi}(q)=i\int d^4 x \, e^{iq\cdot x} 
 \langle T[j^\mu (x)j^\nu(0)] \rangle_{T} \label{eq:cccorrelation}
\end{align}
where we take $j^\mu = \bar{c}\gamma^\mu c$ for $J/\psi$. In this work,
we set both the medium and $c\bar{c}$ at rest so that $\boldsymbol{q}=0$
and $\tilde{\Pi}(q^2)$ becomes the longitudinal part of the polarization
tensor.
For a large $Q^2= -q^2 = -\omega^2  \gg 0$,
the correlation function can be expressed through OPE as
$\tilde{\Pi}(q^2) = \sum_{n}C_n \langle O_n \rangle_{T}$
with $C_n$ and $O_n$ being the perturbative Wilson coefficients and
operators of mass dimension $n$, respectively.
In heavy quark systems such as $J/\psi$, the expansion can be entirely
expressed by gluonic operators
\cite{Shifman_NPB147,Reinders_NPB186,Klingl_PRL82,hayashigaki99:_jpsi}.
Here we assume that the relevant information is contained in the local
operators. Note that, in the deconfined phase, this may have to be
remedied since non-local contributions coming from the non-vanishing
Polyakov loop might become important \cite{medias93}.
The expectation value is taken at finite $T > T_{\text{c}}$ and the
temperature dependence can be imposed only to the expectation value of
the gluonic operator in the case of $m_c \gg T$ and $T^2 \ll Q^2$
\cite{Hatsuda93}. In previous QCD sum rule works,
Hashimoto \textit{et~al.} calculated the Wilson coefficient of the
scalar gluon operator at finite temperature \cite{Hashimoto_ouam}. In
the works of Furnstahl \textit{et~al.} \cite{Furnstahl_PRD42},
scattering contribution was
calculated. However, none of these works included the contribution from the
gluon operator with spin nor the changes of the gluon operators
systematically extracted from the lattice calculations. In contrast, we have
included all the lowest non-vanishing operators, have extracted the full
temperature dependence of the gluon operators from the quenched lattice
data, and have modeled the phenomenological side consistent with the
quenched assumption. Hence, this work marks the first systematic
application of QCD sum rules slightly above $T_{\text{c}}$.

Following Ref.~\cite{Reinders_NPB186}, let us consider the $n$-th moment of
the correlation function \eqref{eq:cccorrelation},
\begin{equation}
  M_n(Q^2) \equiv \left.\frac{1}{n!}\left(\frac{d}{dq^2}\right)^n
 \tilde{\Pi}(q^2)\right|_{q^2 = -Q^2}\label{eq:moment}.
\end{equation}
From the OPE side, up to dimension four, this moment can be expressed as
\begin{gather}
  M_n(Q^2) =
 A_n(\xi)[1+a_n(\xi)\alpha_{\text{s}}+b_n(\xi)\phi_{\text{b}}
 +c_n(\xi)\phi_{\text{c}}]\label{eq:OPE}
\end{gather}
where $\xi = Q^2/4m_c^2$ is a dimensionless scale factor, $A_n$, $a_n$,
$b_n$ and $c_n$ are the Wilson coefficients corresponding to bare loop
diagrams, perturbative radiative correction, scalar gluon condensate and
twist-2 gluon operator, respectively \cite{Klingl_PRL82}. 
Note that
there appears the additional twist-2 contribution when we consider the
medium expectation value. The Wilson
coefficients are listed in Ref.~\cite{Reinders_NPB186,Klingl_PRL82}. 
The explicit forms of $\phi_{\text{b}}$ and $\phi_{\text{c}}$ are
$\phi_{\text{b}}=\frac{4\pi^2}{9(4m_c^2)^2}G_0$ and
$\phi_{\text{c}} = \frac{4\pi^2}{3(4m_c^2)^2}G_2$,
where
$G_0=\left\langle\frac{\alpha_s}{\pi}G_{a\mu\nu}G_a^{\mu\nu}\right\rangle_T
$, and $G_2$ is the twist-2 condensate contribution defined by 
$\left\langle \frac{\alpha_s}{\pi}G_a^{\mu\rho}G_{a\rho}^{\nu}
 \right\rangle_T
 = \left(u^\mu u^\nu -\frac{1}{4}g^{\mu\nu}\right)G_2$ with $u^\mu$
 being the 4-velocity of the medium. These condensates can be determined
 from lattice QCD data as follows.

\begin{figure}[ht]
  \includegraphics[width=3.375in]{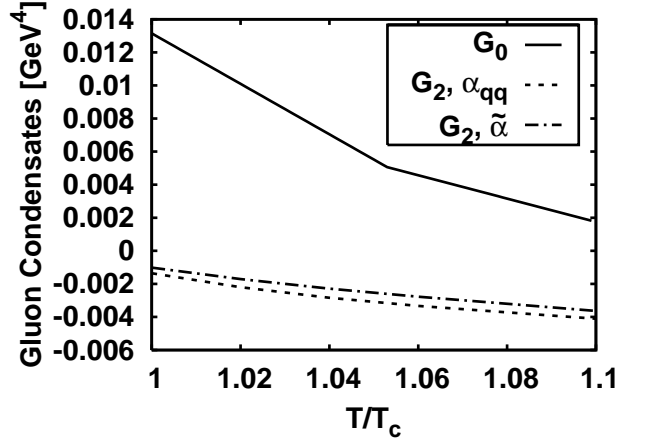}
  \caption{Gluon condensates as a function of $T/T_c$.}
  \label{fig:gc}
 \end{figure}

 The scalar condensate is related to the energy-momentum tensor through the
 trace anomaly. If
 we take 1-loop expression for the beta function of pure SU(3) theory,
 we get $G_0 = G_0^{\text{vac}}- \frac{8}{11} (\varepsilon-3p)$
 where $G_0^{\text{vac}}=(0.35 \text{GeV})^4$ is the value of the gluon
 condensate in vacuum \cite{NoteonG0} and the second term comes from the
 trace anomaly \cite{Miller_hepph2006}. $\varepsilon$ and $p$ are the
 energy density and pressure, respectively. On the
 other hand, the
 twist-2 part can be simply
 related to the energy-momentum tensor of the pure gauge theory as 
 $T^{\alpha\beta} = -G^{\alpha\lambda}_aG^{\beta}_{a\lambda} \quad
 (\alpha\neq\beta)$.
 Hence, recalling that $T^{\alpha\beta} = (\varepsilon+p)u^\alpha
 u^\beta - pg^{\alpha\beta}$, we obtain 
 $G_2 = -\frac{\alpha_s(T)}{\pi}(\varepsilon+p)$.
 The thermodynamic quantities and the effective coupling constant
 $\alpha_s(T)$ are taken from quenched lattice QCD calculation
 \cite{Boyd_NPB469,Kaczmarek_PRD70}. Accounting for the ambiguities of
 $\alpha_s(T)$ in the non-perturbative regime, we adopt two of results in
 Ref.~\cite{Kaczmarek_PRD70} for the temperature dependent coupling
 constant; one is
 determined from the short-distant force and the other is from the
 screening part of the large distant part. The former does not depend on
 temperature at very short distance and takes its maximum values at some
 distance $r_{\text{screen}}$ which decreases with increasing
 temperature. We use the value at this distance and denote it as
 $\alpha_{qq}(T)$ following Ref.~\cite{Kaczmarek_PRD70}. To obtain
 the temperature dependence, we fit the lattice data point (Fig.6 top in
 Ref.~\cite{Kaczmarek_PRD70}) by Bezier
 interpolation, which results in $\alpha_{qq}(T_c)=0.626$. The latter,
 which we denote $\tilde{\alpha}(T)$ as in Ref.~\cite{Kaczmarek_PRD70},
 has a similar value, but error-bars are still too large especially near
 $T_c$. We use the two-loop expression of the running coupling constant
 with a set of parameters given in Ref.~\cite{Kaczmarek_PRD70}. This
 gives $\tilde{\alpha}(T_c)=0.47$.
 The extracted gluon condensates $G_0$ and $G_2$ above but near
 $T_{\text{c}}$ are
 shown in Fig.~\ref{fig:gc}. One thing to note is that, $G_0$ decreases
 to less than half of its vacuum value
 but remains positive near $T_c$ \cite{lee89}.
 We can see that the tensor condensates have non-negligible values near
 $T_c$.

The $n$-th moment in Eq.~\eqref{eq:moment} can also be expressed as
\begin{equation}
 M_n(Q^2) = \int_{0}^{\infty}
 \frac{\rho_{\text{h}}(s)}{(s+Q^2)^{n+1}} ds,\label{eq:dispint}
\end{equation}
where
$\rho_{\text{h}}(s)=\frac{1}{\pi}\tanh\left(\frac{s}{2T}\right)\text{Im}\Pi(s)$
is the phenomenological spectral function which in general includes not
only the pole term but also the continuum and scattering part
\cite{Bochkarev_npb268,Furnstahl_PRD42}. The scattering term, which also
appears in the OPE side and contributes with a delta function at
zero frequency, could be important in the presence of thermal
fermion. However, since we are considering the gluonic medium and have
extracted the condensates from the pure gauge theory, we can
consistently assume that there is no (anti-)quarks which can scatter
with the current.

We can put $\tanh\left(\frac{s}{2T}\right)=1$ due
to much larger
pole mass and continuum threshold than temperature considered
here. Then, hadronic properties
such as mass and width are related to the OPE side
[Eq.~\eqref{eq:OPE}] by putting a phenomenological functional form in
$\text{Im}\Pi(s)$ of the above equation. Here we employ a simple
Breit-Wigner form
\begin{equation}
 \text{Im}\Pi^{\text{pole}}(s) = \frac{f_0 \sqrt{s}
  \Gamma}{(s-m^2)^2+s\Gamma^2},\label{eq:BW}
\end{equation}
to take finite width into account. 
Since we are interested in the lowest lying resonance of the vector
channel, we should choose an appropriate order $n$ so that the moment
[Eq.~\eqref{eq:dispint}] contains information only on the pole term of
the spectral function. Following
Refs.~\cite{Shifman_NPB147,Reinders_NPB186,Klingl_PRL82}, we take the
ratio of the moment $r_n = M_{n-1}/M_n$ and choose moderately large $n$ such
that the contribution from the excited states and continuum  can be
neglected. Therefore, this ratio should approach a constant value at
sufficiently large $n$. However, when $n$ is large, contribution from
higher dimension operators becomes important. At the $n$ value where
$r_n$ is minimum, pole dominance and truncation of the OPE are valid and
the ratio is close to the real asymptotic value, as have been
extensively investigated in the vacuum sum rule for $J/\psi$
\cite{Reinders_NPB186}. In this work we only consider temperature range
in which the same criterion can still be applied and take the minimum
value for $r_n$ to be its asymptotic value.
Hence, in the practical calculation below, we
firstly evaluate the appropriate $n$ for various temperatures by
calculating $r_n|_{\text{OPE}}$. Then, we look for pairs of $m$ and
$\Gamma$ which satisfy the sum rule relation,
$r_n|_{\text{OPE}}=r_n|_{\text{phen.}}$. 
We employ the vegas monte-carlo integration
\cite{press96:_numer_recip_fortr} to treat a very sharp peak in the
dispersion integral [Eq.~\eqref{eq:dispint}] of the phenomenological side.
The relative error in the numerical integration is found to be on the order of
$10^{-6}$ for $m=3$ GeV and $\Gamma=1$ MeV. This numerical accuracy
becomes better as $\Gamma$ increases, as naively expected.
The normalization scale $\xi$ is
chosen as $\xi=1$. We checked that our result does not strongly depend on the
choice of $\xi$ by varying $\xi$ from 0 to 3 \cite{Fullpaper}. Other
parameters of the theory 
are taken from Ref.~\cite{Klingl_PRL82}, $\alpha_s(8m_c^2)=0.21$ and
$m_c=1.24$ GeV. We did not do fine tuning of these parameters to adjust
the vacuum mass of $J/\psi$ since our interest is in the change of mass and
width induced by the hot medium. 
\begin{figure}[ht]
 \includegraphics[width=3.375in]{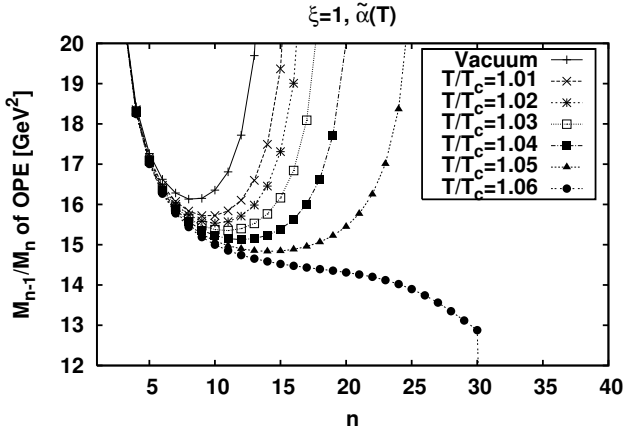}
 \caption{Ratio of the moment from the OPE for various $T/T_c$.}
 \label{fig:ope}
\end{figure}
Figure \ref{fig:ope} displays the ratio of the moment calculated from
Eq.~\eqref{eq:OPE} using $\tilde{\alpha}(T)$ for $G_2$. We can see that
the ratio have a stable point from
vacuum to 1.05$T_c$ but the stability is no longer achieved beyond
1.06$T_c$. Also as seen from the figure, the stable point shifts to
larger $n$ as temperature increases. However, Eq.~\eqref{eq:OPE} becomes
worse as $n$ increases, because the Wilson coefficients increases with $n$
\cite{Reinders_NPB186}. This can be
improved by increasing $\xi$, but the stability holds only up to
1.06$T_c$ even for $\xi=3.0$. If we use $\alpha_{qq}(T)$ for $G_2$, the
stability becomes worse due to its larger value at this temperature
region. In the $\xi=1$ case, there is no stable point for $T=1.05T_c$
with $\alpha_{qq}(T)$. This lack of stability does not necessarily mean
dissociation of $J/\psi$ but shows a breakdown of our approximation.
The reasons for the breakdown are twofold. One is the lack of
convergence of the OPE in Eq.~\eqref{eq:OPE}. This can be improved by
including higher dimensional operators \cite{kim01}. The other is physical one.
Since the non-perturbative part largely decreases above
$T_{\text{c}}$, perturbative contribution will become more important.
Hence, in order to study higher temperature region, we will
need to improve the phenomenological side to be more consistent with the
OPE side, which can be accomplished by a temperature dependent continuum
contribution. Then it will
lead to $n$-independent results for physical parameters until the
$J/\psi$ really dissolves.

\begin{figure}[ht]
 \includegraphics[width=3.375in]{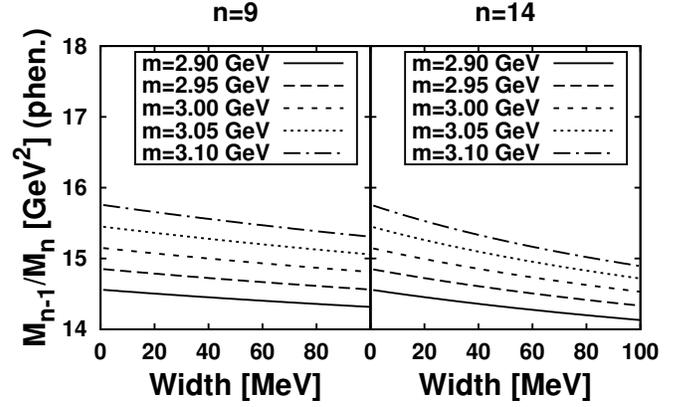}\\
 \caption{The moment ratio of the phenomenological side as a function of
 $\Gamma$ for various masses. Left figure shows $n=9$ case and right
 one shows $n=14$ case.}
 \label{fig:phen}
\end{figure}

Before going to results of $m$ and $\Gamma$, it is useful to see a
feature of the phenomenological side. We depict the moment ratio of the
phenomenological side based on the dispersion integral
\eqref{eq:dispint} in Fig.~\ref{fig:phen}. We choose two $n$ values
which correspond to the stable points at vacuum and $T=1.05T_c$,
respectively. We see that the moment depends on $\Gamma$ very weakly.
For the larger $n$, the dependence becomes slightly stronger. On the
other hand, the moment of the OPE side (Fig.~\ref{fig:ope}) shows about
2 GeV$^2$ decrease from vacuum to $1.05T_c$. Hence, the width must
become very large to achieve 2 GeV$^2$ reduction of the moment ratio if
change of the mass is small. The process to determine the mass and the
width is nothing but evaluating the intersection between the stable
points in Fig.~\ref{fig:ope} and the curves in Fig.~\ref{fig:phen}.
Because of the monotonic behavior of the $r_n$ as a function of $m$ and
$\Gamma$, mass satisfying the sum rule takes its minimum value in the
$\Gamma\rightarrow 0$ limit, \textit{i.e.}, magnitude of the mass shift
becomes the largest if width stays constant. In this limit, mass is
given by a simple relation $m^2 = r_n|_{\text{OPE}}-4m_c^2\xi$
\cite{Reinders_NPB186}. We cannot determine both mass and width only
with the sum rule because it provides only one equation with respect to
two unknown quantities, $m$ and $\Gamma$.  The situation is similar to
light vector meson\cite{leupold98}.
However, we can extract a relation between the
mass shift and width by fixing the mass firstly and then solving
$r_n|_{\text{OPE}}=r_n|_{\text{phen.}}$ for $\Gamma$ because the
monotonic behavior of the $r_n$ guarantees the unique solution.
\begin{figure}[ht]
 \includegraphics[width=3.375in]{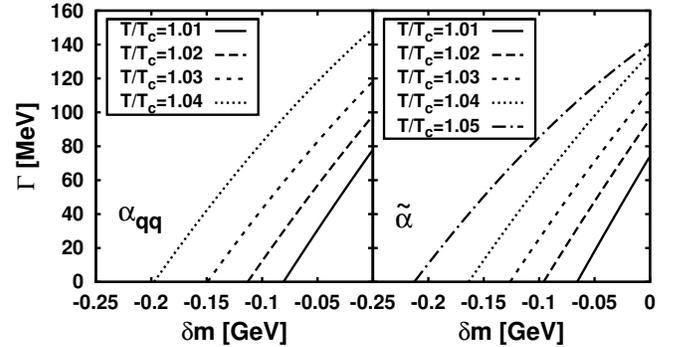}\\
 \caption{Relation between mass shift and width at finite
 temperature. Left figure shows the case of $\alpha_{qq}(T)$ and right
 one shows the case of $\tilde{\alpha}(T)$.}
 \label{fig:dm-g}
\end{figure}

In Fig.~\ref{fig:dm-g}, we display the result of relation between mass
shift $\delta m=m_{\text{medium}}-m_{\text{vacuum}}$ and width which
satisfy the sum rule. The result gives a clear, almost linear relation
between $\Gamma$ and $\delta m$. The difference in $\alpha_s(T)$ appears
as 10-50 MeV difference of the mass shift at $\Gamma=0$ and about 10 MeV
difference of the width at $\delta m=0$. 
\begin{figure}[ht]
 \includegraphics[width=3.375in]{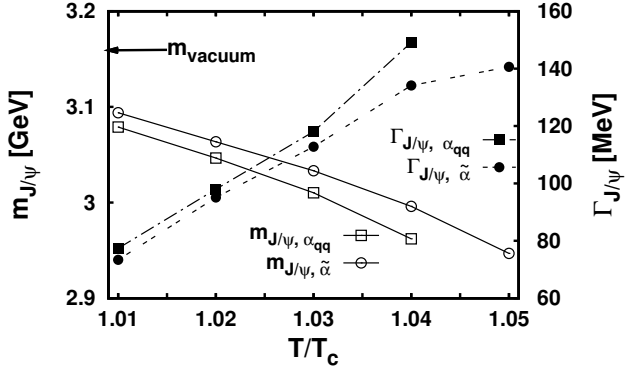}\\
 \caption{Temperature dependence of the $m_{J/\psi}$ (left vertical axis) in
 $\Gamma\rightarrow 0$ limit (max. mass shift) and $\Gamma_{J/\psi}$
 (right vertical axis) in $\delta m\rightarrow 0$ limit (no mass shift). 
 Mass and width are indicated by open symbols and closed ones,
 respectively.}
 \label{fig:T-mg}
\end{figure}

Finally we plot the two extreme cases, $m_{J/\psi}$ for
$\Gamma\rightarrow 0$ and $\Gamma$ for $\delta m \rightarrow 0$, as
a function of temperature in
Fig.~\ref{fig:T-mg}. In both cases, the change
is almost linear with temperature. We stress that, however, these
results are \textsl{extreme} cases.  Our results
show there must be notable change of mass or width, or both of them.
Hence, once either mass or width is estimated by other methods, one can
obtain the other through the relation given in Fig.~\ref{fig:dm-g}. For
example, a pQCD calculation can give thermal width \cite{Lee_Thermalwidth}.
This does not show large ($\sim$ 100 MeV) thermal width near $T_c$. Then the
large mass shift is expected. Such mass shift is in fact expected, as
the sudden reduction of the asymptotic value of the potential just
above $T_c$ seen in a lattice QCD \cite{karsch04}
will inevitably lead to lowering of bound state energy on that potential
\cite{wong05,kim_AdSQCD}. Although the accuracy of the lattice MEM method
is not enough for clear comparison, the spectral
function calculated from a potential model motivated by a full-lattice
QCD shows a large shift of the peak \cite{mocsy07}. A recent full
lattice QCD also shows the shift of the $J/\psi$ peak \cite{aarts07}.

In heavy ion experiments, this mass shift will also change the number of
formed $J/\psi$ according to statistical hadronization. For $T=170$ MeV, 
mass decrease of 100 MeV increases the yield by a factor of 2
since the factor is roughly characterized by $e^{-\delta m/T}$.
Furthermore, a hydrodynamic calculation
shows the lifetime of QGP is $\sim 4-5$ fm/$c$ at 200GeV/$A$ Au+Au
collisions at RHIC \cite{Morita_hydro}. This will be much longer at
LHC. If the width broaden to as large as such lifetime of the plasma,
the mass shift might be detectable.

In summary, we have given the first model-independent analysis of
possible mass shift and
broadening of width of $J/\psi$ on the basis of lattice QCD inputs and
QCD sum rule in the quenched approximation. Although the formalism, OPE
up to dimension 4, is found to
be applicable only to $T\simeq 1.05T_c$, we found that the change of
gluon condensates in the deconfined phase causes notable reduction of
the ratio of the moment
$r_n|_{\text{OPE}}$, which results in mass shift and width broadening.
The mass is found to be almost linear decrease with temperature if the width
remains unchanged while the width linearly increases with temperature in
the case of no mass shift. The maximum values are $\delta m \simeq 200$
MeV and $\Gamma\simeq 140$ MeV at $T=1.05T_c$. Further details including
results for different $\xi$ will be given in a future publication
\cite{Fullpaper}.

\begin{acknowledgments}
 This work is supported by BK21 Program of the Korean Ministry of
 Education. S.~H.~L. was supported by the Korean Research Foundation
 KRF-2006-C00011.
 K.~M. would like to thank the members of high energy physics
 group of Waseda University for providing him their computer facilities
 in which a part of the numerical works was done.
\end{acknowledgments}


\begin{thebibliography}{26}
\bibitem{NoteonG0}It is possible to directly calculate
 $G_0$ in lattice QCD. However, such a procedure is cumbersome as we
 would have to subtract out the perturbative part at $T=0$.
\end{thebibliography}

\end{document}